\documentclass[twocolumn,pre,floats,aps,amsmath,amssymb]{revtex4-1}
\usepackage{graphicx} 
\usepackage{dcolumn} 
\usepackage{bm} 
\usepackage{amssymb} 
\usepackage{amsmath}
\usepackage{txfonts}
\usepackage{color}
\usepackage{ulem}

\begin{document}

\title{Tunable negative permeability in a quantum plasmonic metamaterial}

\author{K. R. McEnery,$^{1,2}$ M. S. Tame,$^{2,3}$ S. A. Maier,$^2$ and M. S. Kim$^1$}

\affiliation{$^1$Quantum Optics and Laser Science Group, Blackett Laboratory, Imperial College London, Prince Consort Road, SW7 2BW, United Kingdom \\
$^2$Experimental Solid State Group, Blackett Laboratory, Imperial College London, Prince Consort Road, SW7 2BW, United Kingdom \\
$^3$School of Chemistry and Physics, University of KwaZulu-Natal, Durban 4001, South Africa}

\date{\today}

\begin{abstract}
We consider the integration of quantum emitters into a negative permeability metamaterial design in order to introduce tunability as well as nonlinear behavior. The unit cell of our metamaterial is a ring of metamolecules, each consisting of a metal nanoparticle and a two-level semiconductor quantum dot (QD). Without the QDs, the ring of the unit cell is known to act as an artificial optical magnetic resonator. By adding the QDs we show that a Fano interference profile is introduced into the magnetic field scattered from the ring. This induced interference is shown to cause an appreciable effect in the collective magnetic resonance of the unit cell. We find that the interference provides a means to tune the response of the negative permeability metamaterial. The exploitation of the QD's inherent nonlinearity is proposed to modulate the metamaterial's magnetic response with a separate control field.
\end{abstract}

\maketitle


\section{Introduction}
\label{sec:intro}

The fields of quantum optics, plasmonics and metamaterials all share the broad purpose of advancing our understanding of light and exploring new ways in which it can be efficiently controlled. The specific focus of each of these fields, however, is markedly different. While quantum optics is the study of light-matter interactions at the quantum level, typically between single atoms and photons~\cite{mandelqo,Mon,milburnqo}, the physical size of the supporting media is of the order of the wavelength of the light or larger, {\it e.g.} in waveguides and cavities. Plasmonics and metamaterials, on the other hand, consider subwavelength media where more compact guiding and confinement structures become possible~\cite{plasmonicsmaier,shalaevmetabook}.  In particular, metamaterials are composed of periodic lattices of identical subwavelength unit cell scatterers, each of which governs completely the electromagnetic properties of the entire bulk material~\cite{shalaevmetabook}. The advancement of metamaterials has led to the development of `left-handed' negative refractive index materials~\cite{veselagonr,Smithrev,Padillarev,McCall} and opened up many exciting applications, such as the super lens~\cite{veselagonr,smithnr,pendrylens}, transformation optics~\cite{trans} and electromagnetic cloaking~\cite{cloak}. 

Recently the fields of quantum optics, plasmonics and metamaterials have increasingly started to overlap with each other. One example is the effort to bring metamaterials from the microwave~\cite{smithnr,Smithrev,Padillarev} to the optical regime~\cite{Soukrev,opticalmeta,opticalmeta2}, which requires the unit cells to be scaled down to the nanoscale to ensure they are subwavelength. Plasmonic structures such as metal nanoparticles (MNPs) are natural electric resonators at optical frequencies~\cite{plasmonicsmaier,kreibigclusters} and so the metamaterial community has increasingly looked towards plasmonics to design optical metamaterials. One of the big challenges, however, has been developing plasmonic magnetic resonators that can be combined with the electric resonators in order to achieve an optical negative refractive index~\cite{salandrino,simovskiring,aluring,Sharing}. A possible route to address this challenge is through the use of unit cells based on quantum plasmonics~\cite{qplasmonics}. Quantum plasmonics combines the fields of quantum optics and plasmonics in order to study the quantum features of surface plasmons, as well as investigating the intense interaction of quantum emitters with localized plasmonic fields at nanostructures. The extra functionality provided by quantum emitter-plasmonic structures in the form of their interaction, tunability and associated interference effects may provide advantages over previous metamaterial designs. In this work, we explore such a scenario, combining quantum optics, plasmonics and metamaterials by investigating the integration of two-level semiconductor quantum dots (QDs) into a negative permeability metamaterial design made up of plasmonic magnetic resonators. The motivation of this work is to bring together phenomena from metamaterials and quantum plasmonics in order to address the problem of achieving a negative refractive index. This work links in with recent studies of transmission array quantum metamaterials~\cite{felbacqmeta}, where theoretical investigations have looked into the possibility of engineering the bulk optical properties of mature quantum systems, such as dielectric-based cavity-atom arrays~\cite{quachqmeta} and superconducting Josephson-qubit lines~\cite{zagoskinqmeta,Shvet,Zuecox,Hutter,Savinov}. Recent experimental work related to these studies has also investigated probing metamaterials in the quantum regime~\cite{zhang}.
\begin{figure*}[t]
\begin{center}
\includegraphics[scale=.6]{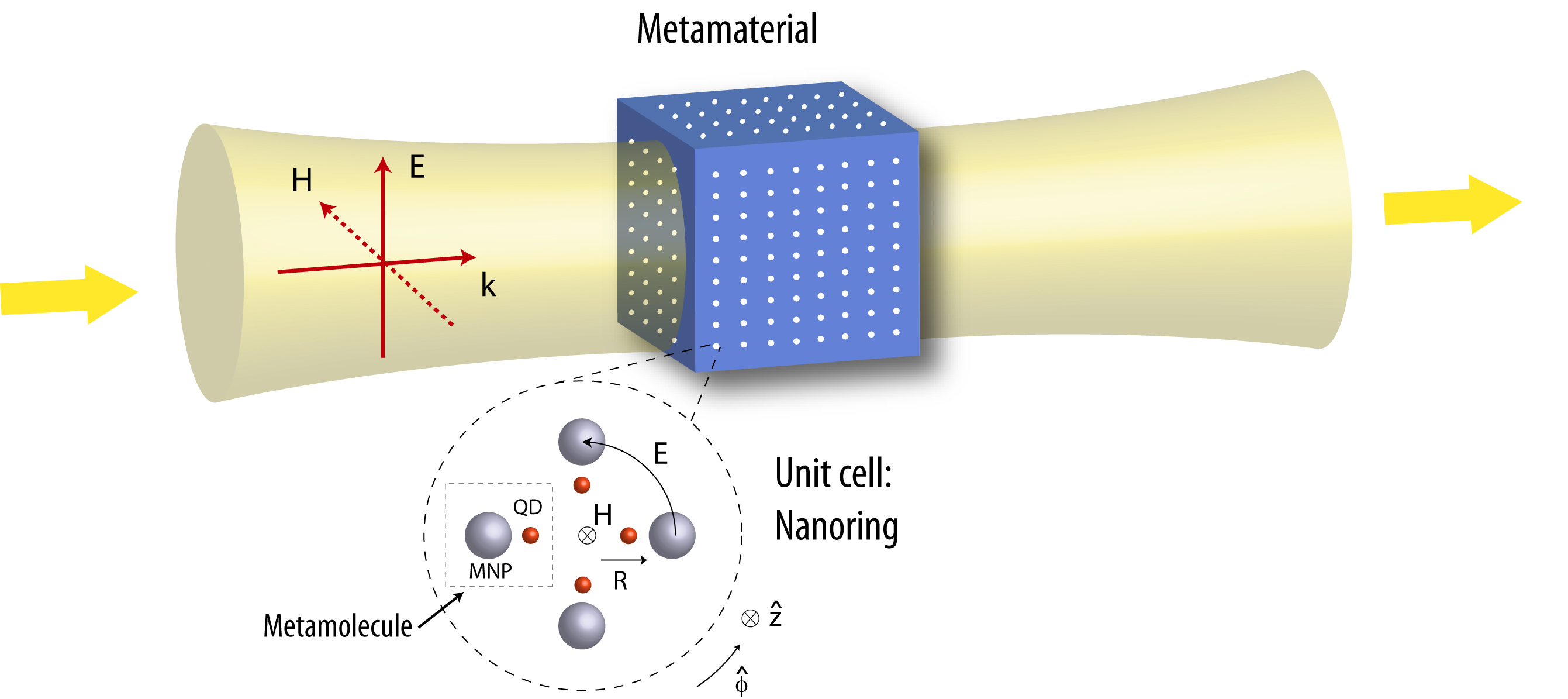}
\caption{\label{fig:1}(Color online) Schematic of the quantum plasmonic metamaterial. This includes a detailed sketch of one of the unit cells consisting of a MNP-QD nanoring. The unit cell has nanoscale dimensions: the radius of the MNP nanoring, QD nanoring, MNPs and QDs are $38$ nm, 6nm, $16$ nm and approximately 1 nm respectively. An arbitrary optical field can be injected into the metamaterial and the figure shows an example of a transverse plane wave at the center of a focused beam. However, in order to isolate the ring's magnetic dipole response, in our study we consider a quasi-static magnetic field $H$ directed along the $\bm{\hat{z}}$ axis, as shown in the unit cell inset. This is a standard approach used to isolate the electric and magnetic response of a metamaterial~\cite{isolate,salandrino}. The MNPs and the QDs are excited by the induced azimuthal E field, as shown in the inset, with no net electric field, thereby isolating the unit cell's magnetic response. The interaction between the QDs and MNP fields at each site, or `metamolecule', cause a Fano profile to appear in the scattered magnetic field of the MNP ring. This induces interference effects in the bulk magnetic response of the metamaterial which enable it to become tunable. The configuration shown is for a material with a magnetic response in the $\bm{\hat{z}}$ direction only and the material is therefore anisotropic. To make an isotropic material with the same response in all directions, a cubic lattice consisting of three orthogonal arrays of nanorings should be used. This can be achieved in a face-centered cubic lattice, where up to four different nanoring orientations can be included in a single unit cell.}
\end{center}
\end{figure*}

In our work we utilize two different types of phenomenon in constructing a quantum plasmonic metamaterial. The first is the formation of an effective optical magnetic dipole from a unit cell of a metamaterial consisting of a coplanar ring of MNPs supporting localized surface plasmon (LSP) modes~\cite{salandrino, simovskiring, aluring}, as shown in Fig.~\ref{fig:1}. The second phenomenon we use is the Fano interference~\cite{fano, Fanoreview}, observed when light scatters from an interacting MNP-QD `metamolecule' system~\cite{zhangqdmnp,savastaqdmnp,waksqdmnp}. By replacing each MNP in the coplanar ring of the unit cell with an MNP-QD metamolecule, we are able to transfer the Fano interference into the magnetic resonance of the ring. When the MNP-QD nanorings are then used as the unit cell of a metamaterial, the Fano interference manifests itself in the material's bulk permeability. We exploit this interference to introduce control over the metamaterial's optical properties, providing tunable and nonlinear responses.

The paper is organized as follows. In Section II, we introduce the physics of the interacting MNP-QD metamolecule and show that a Fano interference profile is present in the linear polarizability of the joint system. In Section III, we discuss the `bare' MNP nanoring magnetic resonator (without QDs) and calculate the bulk permeability of a metamaterial composed of such inclusions. In Section IV, we incorporate the MNP-QD design from Section II into the nanoring and calculate the permeability of our proposed quantum plasmonic metamaterial. We show how the Fano interference effect can be used to tune the metamaterial's properties as well as introduce a nonlinear macroscopic response. We conclude our study in Section V by summarizing our work and outlining future directions of research, including how one might achieve a negative refractive index based on this approach.


\section{The MNP-QD metamolecule}
\label{metamo}

When light of a particular frequency is incident upon a metal nanoparticle of subwavelength size a localized surface plasmon mode (LSP) is excited~\cite{plasmonicsmaier}. The LSP is a non-propagating excitation of the conduction electrons in the MNP and is associated with a large enhancement of the electric field within and in the near field of the MNP. Placing quantum emitters such as atoms, quantum dots and nitrogen vacancy centers within the MNPs strong near field enhances the interaction between them~\cite{qplasmonics}. Typically, light interacts weakly with atoms and other quantum emitters due to the large size mismatch between the spatial extent of the field and the emitter. The MNP provides an interface between the incident light and the emitter, acting much like an antenna. As a result, the coupling frequency between the MNP field and the emitter can become very large. When this coupling frequency dominates all damping rates in the system, the strong-coupling regime is achieved~\cite{milburnqo}. Here, the MNP and emitter coherently exchange energy so rapidly that the two can no longer be considered separately and must be viewed as a joint system. However, due to the large Ohmic and radiative damping associated with the MNP field mode, it is rarely the case that a MNP-emitter system is able to reach this regime, despite the large coupling frequency. Typically the coupling frequency, while lower than the MNP field's damping rate, is much larger than the emission rate of the emitter. This is a regime where the spontaneous emission of the emitter can be enhanced~\cite{purcell}. Furthermore, a Fano interference can occur between the incident field and the excited field in the MNP-emitter system, leading to a characteristic Fano profile in the frequency of the scattered field~\cite{zhangqdmnp, savastaqdmnp, waksqdmnp}. This interference is ubiquitous in wave mechanics and occurs when a discrete system interacts with a continuum~\cite{fano, Fanoreview}. In the present case being considered, the former is the emitter and the latter is the MNP. This particular MNP-emitter type of system has been studied in depth using both a semi-classical~\cite{zhangqdmnp} and a fully quantum mechanical model~\cite{savastaqdmnp,waksqdmnp}. The semi-classical model is perfectly suitable to examine the Fano interference in the weak-driving field limit. However, in the strong-field limit the semi-classical model breaks down and some of the nonlinear behavior predicted is invalidated by quantum noise~\cite{waksqdmnp}. In order to study nonlinear effects in our metamaterial design our model must be able to operate in the strong-field limit. As such, our model is set up from the beginning within a quantum framework.

In this work we focus on producing Fano interferences in a MNP-QD system and exploiting them in our metamaterial design. Thus, we are interested in the explicit optical response of the MNP-QD system. The response of this system to incident light is characterized by a frequency dependent polarizability, $\alpha(\omega)$~\cite{greiner}. For a given incident field amplitude $E_{0}$ it is defined as
\begin{equation}
\alpha(\omega) = \frac{p_{MNP-QD}(\omega)}{E_0} ,
\end{equation}
where $p_{MNP-QD}(\omega)$ is the amplitude of the metamolecule's dipole moment. The polarizability is a complex function where the real part describes the dispersion of the material and the imaginary part its absorption. To derive the MNP-QD system's polarizability, we first define its Hamiltonian, which is given by
\begin{equation}
\hat{H} = \hat{H}_{0} + \hat{H}_{int} + \hat{H}_{drive},
\label{Hdrive0}
\end{equation}
where the individual terms are
\begin{align}
&\hat{H}_0 = \hbar\omega_{0}\hat{a}^{\dagger}\hat{a} + \hbar\omega_{x} \hat{\sigma}^{\dagger}\hat{\sigma}, \\
&\hat{H}_{int} = i\hbar g(\hat{\sigma}\hat{a}^{\dagger}-\hat{\sigma}^{\dagger}\hat{a} ),\\
&\hat{H}_{drive} = - E_{0}\mu(\hat{\sigma}e^{-i\omega t} + \hat{\sigma}^{\dagger}e^{i\omega t})\\ \nonumber
& \quad \quad \quad  - E_{0}(\chi^*\hat{a}e^{-i\omega t} + \chi\hat{a}^{\dagger}e^{i\omega t}).
\end{align}
Here, $\omega_{0}$ and $\omega_{x}$ are the resonance frequencies of the MNP plasmonic field mode and the QD respectively, and $\omega$ is the external driving field frequency. The MNP resonant frequency $\omega_0$ can be derived using the Frohlich condition~\cite{plasmonicsmaier} and by modeling the permittivity of silver using the Drude model, leading to the relation $\omega_0 = \frac{\omega_p}{\sqrt{\epsilon_{\infty} + 2\epsilon_b}}$. The plasma frequency of silver is taken as $\omega_p = 2\pi \times 2175$~THz~\cite{salandrino,aluring,simovskiring} and $\epsilon_{\infty}$ is the ultra-violet permittivity of silver, which is set to $\epsilon_{\infty} = 5$~\cite{simovskiring}. Finally, $\epsilon_b$ is the permittivity of the background material in which the MNP-QD system is embedded.

In Eq.~(\ref{Hdrive0}), the term $\hat{H}_0$ is the free energy Hamiltonian of the MNP and QD, where $\hat{a}^\dag$ ($\hat{a}$) is the creation (annihilation) operator for the MNP plasmonic mode and $\hat{\sigma}^\dag$ ($\hat{\sigma}$) is the raising (lowering) operator for the QD. The term $\hat{H}_{int}$ describes the near-field interaction between the QD and the MNP plasmonic mode, while $\hat{H}_{drive}$ accounts for the driving of the system by an external electric field $E_0$.                                                                                                                                                                                           The coupling of the MNP plasmonic mode to the QD and the driving field are characterized by $g$ and $\chi$ respectively, and $\mu$ is the dipole moment of the QD.

The above Hamiltonians do not account for any losses the system may incur due to interactions with an external environment. The system can lose energy both radiatively to the electromagnetic vacuum, as well as due to Ohmic losses in the metal. These environmental couplings can be modeled as an interaction of the system with a bath of quantized harmonic oscillators~\cite{zhangqdmnp,savastaqdmnp,waksqdmnp}. Treating these interactions with Born-Markov approximations enables the use of a master equation in Lindblad form, $\dot{\hat{\rho}} = \mathcal{\hat{L}}(\hat{\rho})$, which gives a complete description of the system dynamics~\cite{carmichealqo1}. Here, the Lindblad operator acts as follows
\begin{equation}
\mathcal{\hat{L}}(\hat{\rho}) = \frac{i}{\hbar}[\hat{\rho},\hat{H}] + \hat{L}_0 + \hat{L}_x,
\end{equation}
where
\begin{equation}
\hat{L}_j = \frac{\gamma_j}{2}(2\hat{c}_j\hat{\rho}\hat{c}_j^{\dagger} - [\hat{c}_j^{\dagger}\hat{c}_j,\hat{\rho}]_+),
\end{equation}
$j = \{0, x\}$ and $\hat{c}_{0}$ ($\hat{c}_x$) represents $\hat{a}$ ($\hat{\sigma}$). For the decay rates, $\gamma_x$ is the spontaneous emission rate of the QD and $\gamma_0$ accounts for both Ohmic, $\gamma_{nr}$, and radiative damping, $\gamma_r$, of the LSP, where $\gamma_0 = \gamma_{nr} + \gamma_r$. The spontaneous emission rate of the QD is taken as $\gamma_x = 80\times10^{9}$ rad s$^{-1}$~\cite{savastaqdmnp}. On the other hand, the Ohmic damping of the MNP is $\gamma_{nr} = \gamma + \frac{\gamma^3(2\epsilon_b + \epsilon_{\infty})}{\omega_{p}^2}$, where $\gamma$ is the damping frequency of silver which we take as $\gamma = 2.7 \times 10 ^{13}$  rad s$^{-1}$~\cite{salandrino,aluring,simovskiring}. The radiative emission is calculated from a dipole scattering formula, $\gamma_{r} = \frac{2k^3\omega_0r^3\epsilon_b^2}{\epsilon_{\infty}+2\epsilon_b}$~\cite{aluring}, where $k$ is the wavenumber of the light and $\epsilon_0$ is the free space permittivity. Radiative scattering dominates for larger MNPs which are more efficient antennas, while for small MNPs the Ohmic damping dominates as the mean free path of the conduction band electrons decreases~\cite{Brong}.

In order to find the dipole moment of the MNP-QD system and hence its polarizability we need to find the expectation values of the system operators. By working in the Heisenberg picture we can calculate the equations of motion for the expectation values, {\it i.e.} the Maxwell-Bloch (MB) equations, which we express here in a frame rotating with the driving field frequency $\omega$,
\begin{align}
&\langle\dot{\hat{a}}\rangle = -(i\Delta_0  + \frac{\gamma_0}{2})\langle\hat{a}\rangle + g\langle\hat{\sigma}\rangle + \frac{i\chi E_0}{\hbar}, \label{MB1} \\
&\langle\dot{\hat{\sigma}}\rangle = -(i\Delta_x  + \frac{\gamma_x}{2})\langle\hat{\sigma}\rangle - g\langle\hat{a}\rangle +2g\langle\hat{a}\hat{\sigma}^{\dagger}\hat{\sigma}\rangle \label{MB2} \\
& \qquad ~~  + \frac{i\mu E_0}{\hbar}(1-2\langle\hat{\sigma}^{\dagger}\hat{\sigma}\rangle), \nonumber
\end{align}
where $\Delta_{0(x)} = (\omega_{0(x)} -\omega)$. In the general case, the above equations are difficult to solve as they are not in a closed form and thus form an infinite hierarchy of equations~\cite{armenthesis}. However, we can make approximations that transform the equations into more amenable semi-classical equations. This can be done by making the assumption that the QD and the MNP field are separate systems and factoring the term $\langle\hat{a}\hat{\sigma}^{\dagger}\hat{\sigma}\rangle$ into its light and matter components. This is a reasonable assumption when considering that the large damping of the MNP field inhibits coherent interactions~\cite{waksqdmnp}. The MB equations can then be simplified further by assuming a weak driving field. In this case, the excited state population of the QD, $\langle\hat{\sigma}^{\dagger}\hat{\sigma}\rangle$, is taken to be negligible~\cite{zhangqdmnp,savastaqdmnp,waksqdmnp}. 

The MNP-QD system described above is a driven-dissipative one and therefore we are interested in calculating the polarizability when the system reaches a non-equilibrium steady state (NESS), {\it i.e.} when the system operators $\dot{\hat{O}} = 0$. Using the above simplifications the NESS value of the MNP plasmonic field annihilation operator can be found to be
\begin{equation}
\langle\hat{a}\rangle = \frac{g\langle\hat{\sigma}\rangle}{i\Delta_0+ \frac{\gamma_0}{2}} + \frac{i\chi E_0}{\hbar(i\Delta_0 + \frac{\gamma_0}{2})}. \label{expmnp}
\end{equation}

To ensure the above quantum framework describes the physics of the system we compare its results to those predicted by classical theory. In this way we can calculate the parameters $g$ and $\chi$~\cite{savastaqdmnp},
\begin{align}
&g = \frac{S\mu}{d^3}\sqrt{\frac{3\eta r^3}{4\pi\epsilon_0\hbar}}, \label{g}, \\
&\chi = -i\epsilon_b\sqrt{12\eta\epsilon_0\pi\hbar r^3}. \label{chi2}
\end{align}
This derivation is expanded upon in Appendix.~\ref{appA}. Here, the distance between the MNP and the QD is given by $d$ and the MNP  radius is $r$. The dipole moment of the QD is $\mu = er_0$, where the dipole moment radius is $r_0 = 0.9$ nm (corresponding to 43.22 Debye)~\cite{savastaqdmnp}. The background permittivity is $\epsilon_b = 2.2$, the parameter $S = 2~(-1)$ is set for the external driving field being parallel (perpendicular) to the MNP-QD separation vector and $\eta = \frac{(\gamma^2(2\epsilon_b + \epsilon_{\infty}) + \omega_p^2)^2}{2(2\epsilon_b+\epsilon_{\infty})^{\frac{3}{2}}\omega^3_p}$.
\begin{figure}[t]
\begin{center}
\includegraphics[scale=0.4]{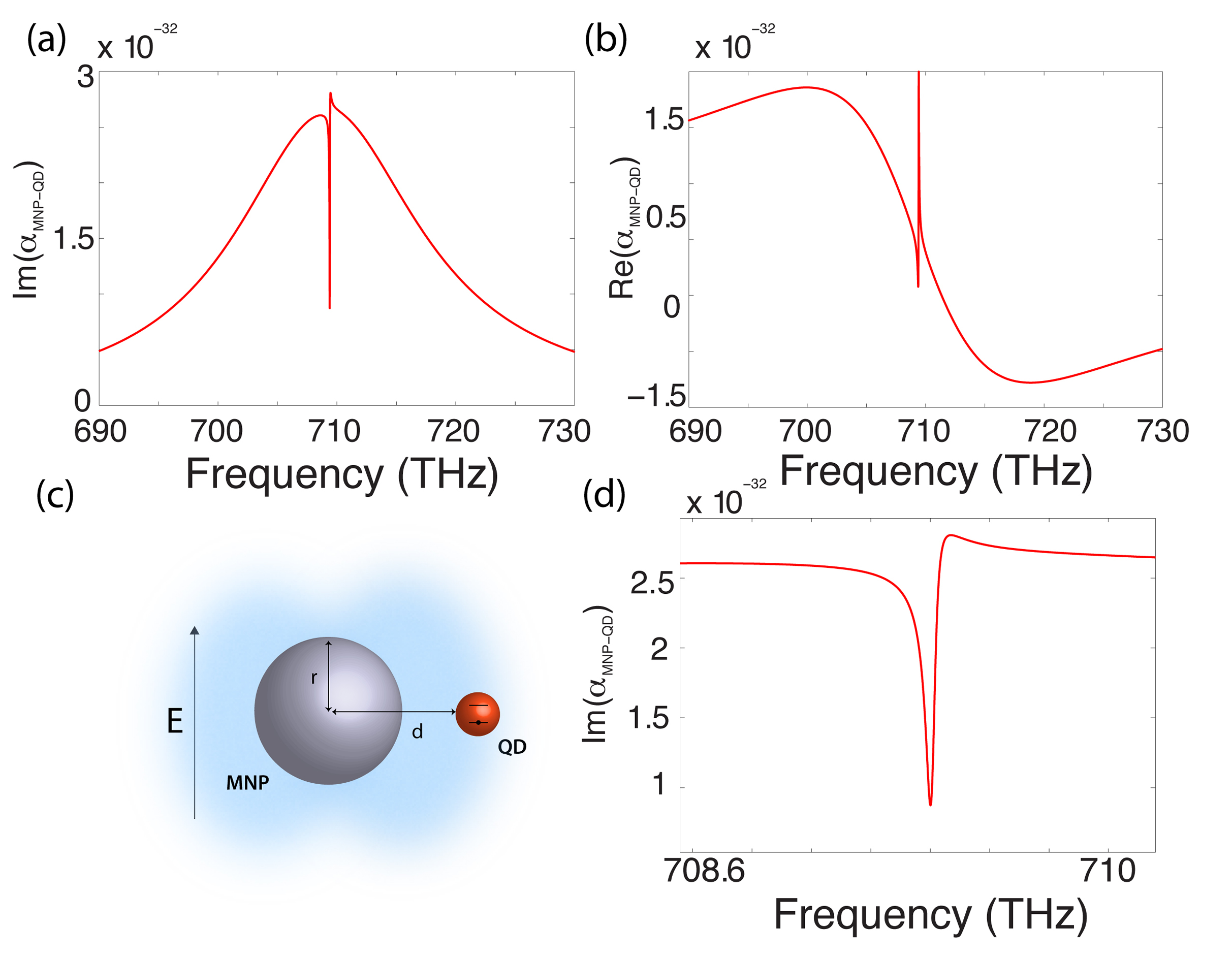}
\caption{\label{fig:2}(Color online) Imaginary (a) and real (b) parts of the polarizability of a MNP-QD metamolecule, whose individual dipole radii are $16$ nm and $0.9$ nm respectively, and whose separation distance is $32$ nm. The MNP and QD are resonant in this example and they couple transversely ($S=-1$). The system is encased in a background medium of permittivity $\epsilon_b = 2.2$ and is driven weakly, $E_0\mu = 0.0001$ meV, where $\frac{E_0\mu}{\hbar}$ is the coupling frequency between the QD and the driving field. In (c) a sketch of the system shows that the MNP and QD are transversely coupled. In (d) we show the imaginary polarizability around the resonance frequency, highlighting the Fano dip.}
\end{center}
\end{figure}
In Appendix~\ref{appA}, the dipole moment of the MNP field is shown to be
\begin{equation}
p_{_{MNP}} = \chi^{*}\langle\hat{a}\rangle \label{mnpdipole}
\end{equation}
From this, an expression for the polarizability of the joint MNP-QD metamolecule system can be derived as
\begin{align}
\alpha(\omega) = \frac{\chi^*\langle\hat{a}\rangle + \mu\langle\hat{\sigma}\rangle}{E_0}.
\end{align}
Then, by solving the MB equations, Eqs.~(\ref{MB1}) and (\ref{MB2}), in the steady state, an analytic expression for the system's polarizability is found to be
\begin{eqnarray}
\alpha(\omega) &=&  \frac{i\mu^{2}}{\hbar(i\Delta_x  + \frac{\gamma_x}{2} + \frac{g^2}{i\Delta_0 + \frac{\gamma_0}{2}})} +  \frac{i|\chi|^{2}}{\hbar(i\Delta_0 + \frac{\gamma_0}{2} + \frac{g^2}{i\Delta_x + \frac{\gamma_x}{2}})} \nonumber \\
& &+ \frac{i\mu g(\chi^{*}-\chi)}{\hbar((i\Delta_x +\frac{\gamma_x}{2})(i\Delta_0 +\frac{\gamma_0}{2}) + g^{2})}.
\end{eqnarray}

In Fig.~\ref{fig:2}~(a) and (b) we show the imaginary and real parts of the metamolecule's polarizability for a range of driving field frequencies, with the resonant frequency of the MNP and QD set to be equal. One can clearly see the Fano interference profile due to the MNP-QD interaction. In addition, looking closer at the imaginary part of the polarizability in Fig.~\ref{fig:2}~(d), one can see that the interference suppresses light absorption at the resonance frequency. The real part of the polarizability is used to calculate the dispersion of the MNP-QD molecule. The frequency regions either side of the resonance are governed by anomalous dispersion where the polarizability decreases with increasing frequency, whereas at resonance there is a sharp increase in polarizability with increasing frequency, {\it i.e.} normal dispersion. This effect is also seen in EIT systems where it is responsible for slow light propagation~\cite{EIT}.

The Fano interference effect can be amplified by increasing the MNP-QD coupling frequency. In the above example, this coupling frequency is quite strong due to the small separation distance chosen, $g/\omega_0 =5\times10^{-4}$. However, if the QD is placed too close to the MNP, then higher order multipoles are excited in the MNP and the dipole approximation breaks down~\cite{multipole}. This should be avoided if we wish to use this scatterer in a metamaterial design using dipole formulae. The QD must also be placed further than 1 nm from the MNP surface in order to avoid electron tunneling~\cite{Zuloaga2009}. We have placed the QD at a distance of $2r$ which is sufficient for higher order multipoles to be negligible~\cite{multipole} as well as to avoid tunneling effects~\cite{Zuloaga2009}.


\section{The MNP nanoring}
\label{mnpring}

We now consider a ring of MNPs in a specific configuration that has recently been studied for its application as a magnetic resonator in the visible regime~\cite{salandrino, simovskiring, aluring,Sharing}. Our goal is to describe this `bare' system quantum mechanically so that we can incorporate the metamolecule from the previous section, and thereby investigate the tunability and nonlinear response provided by the QD emitter.

In order to calculate the permittivity or permeability of a metamaterial, it is common practice to isolate either its electric or magnetic response with a particular type of incident field.  Despite the permittivity and permeability being calculated using a special type of excitation method, these characteristic functions of the metamaterial approximate well the response of the metamaterial to an arbitrary form of incident field~\cite{isolate,salandrino}. Thus in our work we will concentrate on isolating the magnetic response of the ring. To achieve this, we direct a high frequency magnetic field along the ring's normal axis~\cite{aluring}, as shown in Fig.~\ref{fig:1}. The MNPs thus feel the following electric field 
\begin{equation}
\bm{E_{0}} = \frac{i\omega\mu_{0}RH_{0}}{2}\bm{\hat{\phi}},
\end{equation}
induced by the time varying incident magnetic field $\bm{H} = H_{0}e^{-i\omega t}\bm{\hat{z}}$. Here, $R$ is the radius of the ring, $\omega$ is the frequency of the driving field, and $\bm{\hat{\phi}}$, and $\bm{\hat{z}}$ are unit vectors in the cylindrical coordinate system $(\bm{\hat{R}},\bm{\hat{\phi}},\bm{\hat{z}})$. The displacement field induced in each MNP is also directed along the azimuthal direction. Due to this symmetry, there is no net electrical response and we are therefore able to isolate the magnetic response of the system. A circular displacement field current is set up which acts as a magnetic dipole, whose magnitude is given by~\cite{salandrino}
\begin{equation}
m = \frac{-i\omega p_{MNP}NR}{2}, \label{magdipole}
\end{equation}
where $N$ is the number of electric dipoles in the ring and $p_{MNP}$ is the dipole moment of a single MNP. In Fig.~\ref{fig:1} the ring configuration we consider is shown for a material with a magnetic response in the $\bm{\hat{z}}$ direction only. Thus, the material is anisotropic. To make an isotropic material with the same response in all directions, a cubic lattice consisting of three orthogonal arrays of nanorings should be used. This can be achieved in a face-centered cubic lattice, where up to four different nanoring orientations can be included in a single unit cell~\cite{salandrino}. Furthermore, in our calculations we concentrate on the case of $N$=4 as it is the minimum number of MNPs in the ring such that the magnetic dipolar response dominates higher order multipoles~\cite{aluring}. This is essential for the validity of characterizing the metamaterial's magnetic response with the permeability parameter, $\mu_{eff}$, which we now derive~\cite{simovskiIOP2011}.

We start by calculating the dipole moment of one of the MNP inclusions using a quantum framework. Although strictly speaking this approach is not required as the process is essentially classical, we set up the quantum formalism now so that it can be used when we integrate MNP-QD metamolecules into the ring in the next section. Thus the equations of motion derived using this framework are also valid in the classical regime. The Hamiltonian of the bare system is as follows
\begin{align}
&\hat{H} = \hat{H}_{0} + \hat{H}_{int} + \hat{H}_{drive},
\end{align}
where the individual terms are
\begin{align}
&\hat{H}_0 =\displaystyle\sum\limits_{n=0}^{N-1} \hbar\omega_{0}\hat{a}_n^{\dagger}\hat{a}_n, \\
&\hat{H}_{int} =\displaystyle\sum\limits_{n,m=0}^{N-1} \hbar J_{nm}(\hat{a}_n^{\dagger}\hat{a}_m +\hat{a}_m^{\dagger}\hat{a}_n)~~~n\not = m\\
&\hat{H}_{drive} = - E_{0}\displaystyle\sum\limits_{n=0}^{N-1}(\chi^*\hat{a}_ne^{-i\omega t} +\chi\hat{a}_n^\dag e^{i\omega t}).
\end{align}
Here, the inter-MNP coupling frequency is given by $J_{nm}$, which for nearest neighbor coupling we denote as $J_1$ and for next-nearest neighbor coupling as $J_2$. The expressions for each are derived in Appendix~\ref{appB} and given by
\begin{align}
J_{1(2)} = -12\pi\epsilon_0\epsilon_b^2r^3\eta Q_{1(2)},
\end{align}
where $Q_{1(2)}$ is the nearest (next-nearest) neighbor scalar interaction term between the MNPs which includes both near field and radiative interactions~\cite{aluring}. Explicitly we have
\begin{align}
&Q_{1} = \frac{e^{i\sqrt{2}kR}}{16\sqrt{2}\pi\epsilon_{0}\epsilon_bR^5}(-2k^2R^4 + 3R^2(1-ik\sqrt{2}R)),\\
&Q_{2} = \frac{e^{i2kR}}{128\pi\epsilon_{0}\epsilon_bR^5}(-16k^2R^4 + 4R^2(1-i2kR)),
\end{align}
where $k$ is the wave vector of the light, $k = \omega\sqrt{\mu_0\mu_b\epsilon_0\epsilon_b}$, where $\epsilon_0~(\mu_0)$ is the free space permittivity (permeability) and $\epsilon_b~(\mu_b)$ is the relative permittivity (permeability) of the background medium.
\begin{figure}[t!]
\begin{center}
\includegraphics[scale=0.95]{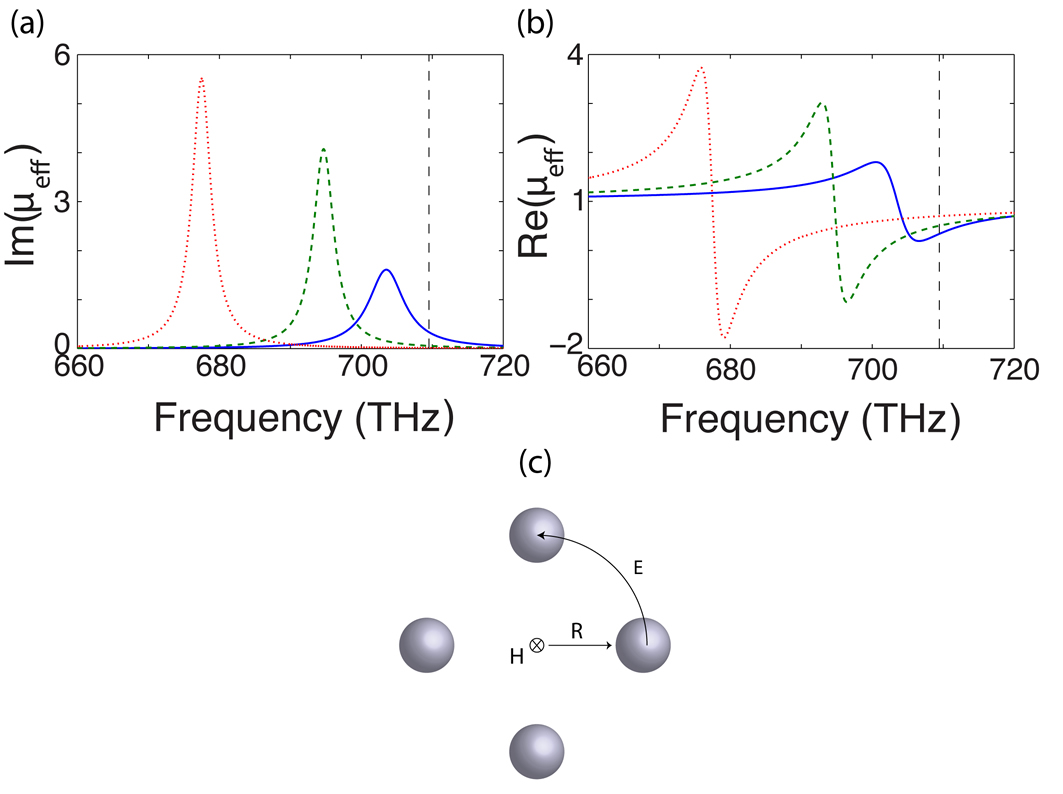}
\caption{\label{fig:4}(Color online) Imaginary (a) and real (b) parts of the effective permeability of an MNP nanoring metamaterial. The ring radius, $R$, and MNP radius, $r$, are $38$nm and $16$nm respectively. The MNPs are encased in a material with permittivity $\epsilon_b = 2.2$ and permeability $\mu_b = 1$. We plot the effective permeability for three values of $N$ in each nanoring, $N$ = 2 (blue line), $N$ = 3 (green dashed line) and $N$ = 4 (red dotted line), where $N_d$ = (96nm)$^{-3}$. The vertical dashed line corresponds to the electric resonance of a single MNP. In (c) we show a sketch of the nanoring system.}
\end{center}
\end{figure}

In the Heisenberg picture, using the full system Hamiltonian, the equation of motion for the expectation value of the annihilation operator of each MNP field mode, $\langle\hat{a}_n\rangle$, can be found and for $N=4$ they can be written as
\begin{align}
\langle\dot{\hat{a}}_n\rangle & = -(i\Delta_0 + \frac{\gamma_0}{2})\langle\hat{a}_n\rangle - iJ_1\langle\hat{a}_{n+1}\rangle - iJ_1\langle\hat{a}_{n-1}\rangle \label{aringexp} \\ 
& ~~~~ -iJ_2\langle\hat{a}_{n+2}\rangle +\frac{i\chi E_0}{\hbar}, \nonumber
\end{align}
where the indices are written in modulo 4. This set of coupled equations can be solved in a straightforward manner in the steady state as the system's symmetry means that the expectation value of each dipole is the same. Using Eq.~(\ref{chi2}) and Eq.~(\ref{mnpdipole}) the dipole moment of a single MNP can be written as
\begin{equation}
p_{_{MNP}} =\bigg(\frac{-|\chi|^2\omega\mu_{0}RH_{0}}{2\hbar(i\Delta_0 + \frac{\gamma_0}{2})}\bigg)\bigg(1 + \frac{i(2J_1 + J_2)}{i\Delta_0 + \frac{\gamma_0}{2}}\bigg)^{-1}. \label{MNPdip}
\end{equation}
Then, using Eq.~(\ref{magdipole}) we can calculate the magnetic polarizability of a single nanoring, $\alpha_m = \frac{m}{H_0}$. The effective permeability of the macroscopic composite system (metamaterial) can be calculated using the Maxwell-Garnett mixing formula~\cite{shalaevmetabook,simovskiIOP2011}, 
\begin{equation}
\mu_{eff} = 1 + \frac{1}{N_{d}^{-1}(\alpha_{m}^{-1} + i\frac{k^3}{6\pi}) - \frac{1}{3}}, \label{mueff}
\end{equation} 
where $N_{d}$ is the volume concentration of nanorings in the composite system. The imaginary term in the denominator is only necessary when the rings are part of a regular three dimensional array. In this case, the radiative damping of the magnetic dipole is cancelled out~\cite{aluring}.

In Fig.~\ref{fig:4} we plot the effective permeability of a metamaterial with nanorings that have 2, 3 and 4 MNP inclusions. As mentioned earlier, only for $N=4$ is the effective permeability physically meaningful, however, it is informative to plot $N=2$ and 3, as they show the effect the inter-MNP coupling has on red-shifting the nanoring's magnetic resonance from the electric resonance of a single MNP (vertical dashed line). In both Fig.~\ref{fig:4}~(a) and (b) one can see the material's resonance properties, including negative real values in panel (b). For the MNP nanoring we use the parameters of Ref.~\cite{salandrino}. Apart from a slightly smaller volume concentration of $N_d$ = (96nm)$^{-3}$, all other parameters are the same.


\section{The MNP-QD nanoring}
\label{mnpqdring}

\begin{figure}[t!]
\begin{center}
\includegraphics[scale=0.95]{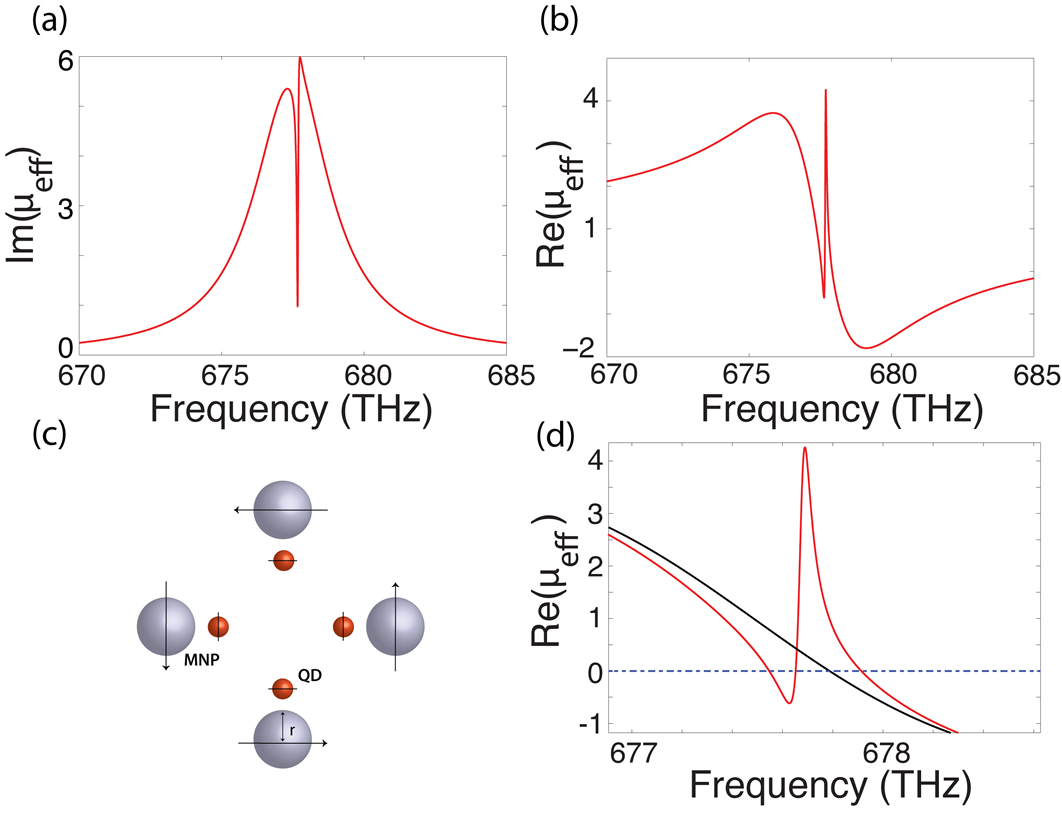}
\caption{\label{fig4F}(Color online) Imaginary (a) and real (b) parts of $\mu_{eff}$ for a QD-MNP nanoring metamaterial. The parameters of the system are chosen to be the same as in Figs.~\ref{fig:2} and~\ref{fig:4}, with a MNP-QD detuning $\Delta = (\omega_0 - \omega_x) = 0.195 \times 10^{15}$ rad s$^{-1}$. In (c) a sketch of the system is shown, where the unit cell now consists of two interacting rings, the MNP ring and the QD ring. The arrows on the MNPs show the electric field direction at each MNP and their direction also represents their adjacent QD's dipole orientation. In (d) we examine the difference between Re($\mu_{eff}$) with (red) and without (black) QDs in the nanoring.}
\end{center}
\end{figure}

We now take the nanoring design from the previous section and replace each MNP with the MNP-QD metamolecule from Section~\ref{metamo}, as shown in Fig.~\ref{fig4F}~(c). In this case there are two magnetic dipoles excited by the incident magnetic field; one set up by the ring of MNPs and the other by the QD ring. We use Eq.~(\ref{mueff}) again to calculate the effective permeability of a metamaterial composed of these MNP-QD nanorings. However, in this case we must deal with two magnetic dipole excitations, as well as taking into account MNP-QD interactions. The Hamiltonian for the system is
\begin{align}
&\hat{H} = \hat{H}_{0} + \hat{H}_{int}+ \hat{H}_{drive},
\end{align}
where the individual terms are
\begin{align} 
&\hat{H}_0 =\displaystyle\sum\limits_{n=0}^{N-1} \hbar\omega_{0}\hat{a}_n^{\dagger}\hat{a}_n + \displaystyle\sum\limits_{n=0}^{N-1}\hbar\omega_{x} \hat{\sigma}_n^{\dagger}\hat{\sigma}_n, \\
&\hat{H}_{int} =\displaystyle\sum\limits_{n,m=0}^{N-1} \hbar J_{nm}(\hat{a}_n^{\dagger}\hat{a}_m + \hat{a}_m^{\dagger}\hat{a}_n)~~~n\not = m \\
&\quad \quad +\displaystyle\sum\limits_{n,m=0}^{N-1} \hbar I_{nm}(\hat{\sigma}_n^{\dagger}\hat{\sigma}_m + \hat{\sigma}_m^{\dagger}\hat{\sigma}_n)~~~n\not = m  \nonumber
\\
&\quad \quad + \displaystyle\sum\limits_{n,m=0}^{N-1} i\hbar g_{nm}(\hat{a}_n^{\dagger}\hat{\sigma}_{m} + \hat{a}_n\hat{\sigma}_{m}^{\dagger}),\nonumber \\
&\hat{H}_{drive} = - E_{0}\displaystyle\sum\limits_{n=0}^{N-1}(\chi^*\hat{a}_ne^{-i\omega t} + \chi\hat{a}_n^\dag e^{i\omega t})\\ \nonumber &\quad \quad \quad - E_{0}\mu\displaystyle\sum\limits_{n=0}^{N-1}(\hat{\sigma}_ne^{-i\omega t} + \hat{\sigma}_n^\dag e^{i\omega t}).
\end{align}

First we will calculate the effective permeability in the weak-field limit. In this case we can derive steady state MB matrix equations accounting for each site
\begin{align}
{\bf A}\bar{a} = {\bf B}\bar{\sigma} + \bar{c} \\
{\bf D}\bar{\sigma} = -{\bf B}\bar{a} + \bar{e}
\end{align}
Where $\bar{a}$ and $\bar{\sigma}$ are vectors which represent the expectation values for $\hat{a}$ and $\hat{\sigma}$ at each site in the ring, given as
\begin{equation}
\bar{a} = \begin{pmatrix}\langle\hat{a}_1\rangle\\\langle\hat{a}_2\rangle\\\langle\hat{a}_3\rangle\\\langle\hat{a}_4\rangle\end{pmatrix} , \bar{\sigma} = \begin{pmatrix}\langle\hat{\sigma}_1\rangle\\\langle\hat{\sigma}_2\rangle\\\langle\hat{\sigma}_3\rangle\\\langle\hat{\sigma}_4\rangle\end{pmatrix}.
\end{equation}
The matrix ${\bf A}$ represents the MNP-MNP interactions in the nanoring, given as
\begin{equation}
{\bf A} = \begin{pmatrix}i\Delta_0 + \frac{\gamma_0}{2}&iJ_1&iJ_2&iJ_1\\iJ_1&i\Delta_0 + \frac{\gamma_0}{2}&iJ_1&iJ_2\\iJ_2&iJ_1&i\Delta_0 + \frac{\gamma_0}{2}&iJ_1\\iJ_1&iJ_2&iJ_1&i\Delta_0 + \frac{\gamma_0}{2}\end{pmatrix},
\end{equation}
where the MNP-MNP coupling frequency $J_n$ was defined in the previous section. The matrix ${\bf D}$ represents the QD-QD interactions in the nanoring, given as
\begin{equation}
{\bf D} = \begin{pmatrix}i\Delta_x + \frac{\gamma_x}{2}&iI_1&iI_2&iI_1\\iI_1&i\Delta_x + \frac{\gamma_x}{2}&iI_1&iI_2\\iI_2&iI_1&i\Delta_x + \frac{\gamma_x}{2}&iI_1\\iI_1&iI_2&iI_1&i\Delta_x + \frac{\gamma_x}{2}\end{pmatrix},
\end{equation}
where $I_1$ and $I_2$ are the nearest neighbor and next-nearest neighbor coupling frequencies, given by
\begin{equation}
I_{1(2)} = \frac{\mu^2}{\hbar}Q_{1(2)}.
\end{equation}
The matrix ${\bf B} $ is the MNP-QD coupling matrix, given as
\begin{equation}
{\bf B} = \begin{pmatrix}g_1&0&-g_2&0\\0&g_1&0&-g_2\\-g_2&0&g_1&0\\0&-g_2&0&g_1\end{pmatrix}.
\end{equation}
Here, the coupling frequency $g_1$ is for same-site MNP-QD interactions, while $g_2$ is for an MNP coupling with its next-nearest QD neighbor. From Fig~\ref{fig4F}~(c) we can see that the same-site and next-nearest neighbor interactions are transverse (S = -1). Due to the azimuthal external electric field exciting the ring, the same-site QD and next-nearest neighbour QD relative to each MNP are driven in opposite directions. As such they are out of phase, this is represented by the minus signs in the matrix. We find that the next-nearest neighbour QD works to reduce the influence of the same-site QD on each MNP. Fortunately due to the stronger same-site interaction frequency the Fano interference still occurs. Similarly both nearest neighbour QDs relative to an MNP are also out of phase, however in this case their interaction frequency is the same and their effect on the MNP is cancelled out. Finally, the vectors $\bar{c}$ and $\bar{e}$ represent the external driving of the MNPs and the QDs by the induced electric field,
\begin{equation}
 \bar{c} = \frac{\chi\omega\mu_0R_1H_0}{2\hbar }\begin{pmatrix}1\\1\\1\\1\end{pmatrix},
\bar{e} = \frac{\mu\omega\mu_0R_2H_0}{2\hbar }\begin{pmatrix}1\\1\\1\\1\end{pmatrix},
\end{equation}
where we have taken into account the differing radii of the MNP ($R_1$) and QD ($R_2$) nanorings. We can solve these equations to calculate the dipole moment of each MNP and QD within the nanoring,
\begin{align}
P_{MNP} = \chi^*\bar{a}_1 = \chi^*({\bf A} + {\bf B} ({\bf D}^{-1}){\bf B})^{-1}({\bf B}({\bf D}^{-1})\bar{e} + \bar{c}),\\
P_{QD} = \mu\bar{\sigma}_1 = \mu({\bf D} + {\bf B} ({\bf A}^{-1}){\bf B})^{-1}({\bf B}({\bf A}^{-1})\bar{c} + \bar{e}).
\end{align}
Due to the symmetry of the system the dipole moment is the same on each site for the MNPs and also for the QDs. Following the procedure in Section III we calculate the magnetic dipole of both the MNP and QD rings. The total magnetic polarizability of the MNP-QD nanoring can be found using the relation $\alpha_m = \frac{m_{_{MNP}} + m_{_{QD}}}{H_0}$ and Eq.~(\ref{mueff}) can be used to calculate the effective permeability of a metamaterial made from the MNP-QD nanorings.

The effective permeability ($\mu_{eff}$) of the metamaterial is shown in Fig.~\ref{fig4F}~(a) and (b). Due to the red-shift of the magnetic resonance of the MNP ring, as shown in Section~\ref{mnpring} ({\it c.f.} Fig.~\ref{fig:4}), the QD resonances have been red-shifted in order to ensure Fano interference. The MNP ring has a radius of $R_1 = 38$~nm, while the QD ring has a radius of $R_2 = 6$~nm. Thus the same site MNP-QD separation is $d = 32$~nm, as used for the MNP-QD molecule in Section I. However, we can see that the Fano interference present in the effective permeability is more prominent than that observed in the polarizability of an individual MNP-QD molecule ({\it c.f.} Fig.~\ref{fig:2}). This is due to the multiple QDs in the nanoring interacting with each MNP either directly or mediated through MNP-MNP interactions. In Fig.~\ref{fig4F}~(d) we show how the introduction of the QDs in the nanoring design changes the real part of the permeability from positive to negative through the Fano interference. While the magnitude of the Fano-affected Re($\mu_{eff}$), when negative, is not very large, as the magnitude is dependent on the strength of the magnetic resonator, the strength of the resonance can be amplified by increasing the number of sites in the ring~\cite{aluring}.
\begin{figure}[t]
\begin{center}
\includegraphics[scale=0.7]{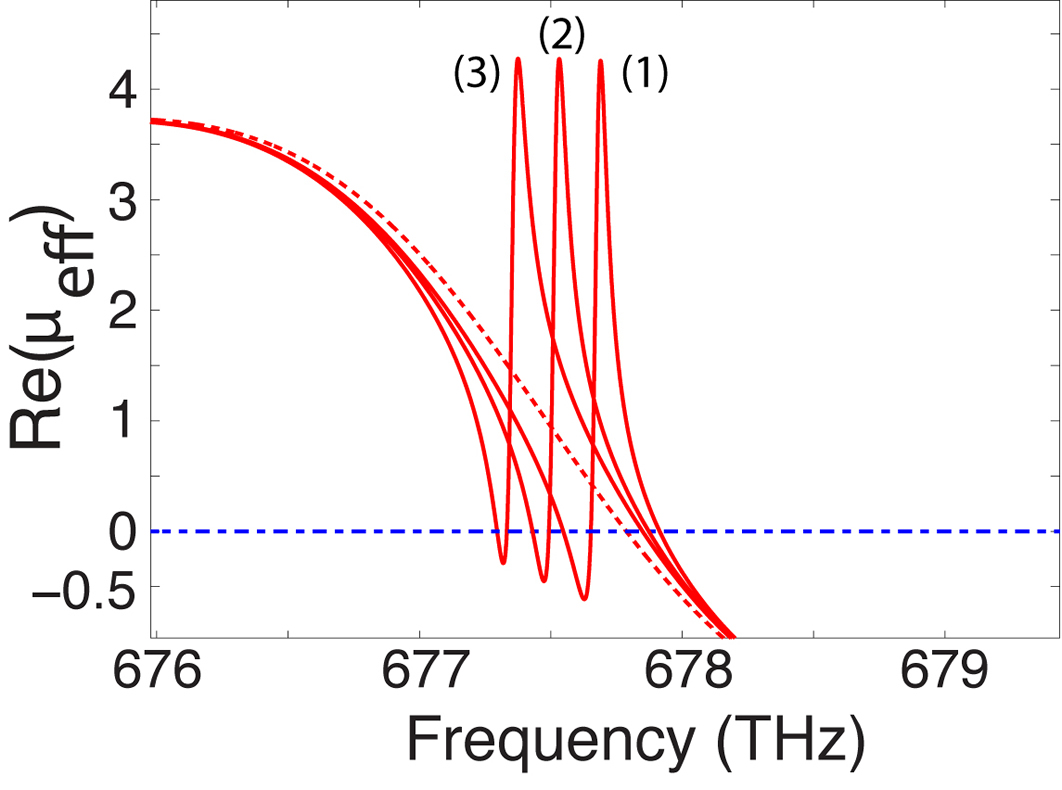}
\caption{\label{fig:5}(Color online) Tunability of the MNP-QD nanoring metamaterial. The real part of $\mu_{eff}$ for different MNP-QD detunings. We plot: (1) $\Delta = 0.195$, (2) $\Delta =0.196$, and (3) $\Delta =0.197$. All detunings are in units of $10^{15}$~rad s$^{-1}$. The dashed line shows the bare MNP nanoring in the absence of QDs.}
\label{fig5}
\end{center}
\end{figure}

From the above analysis, one can see that the integration of QDs in the MNP nanoring has transferred the Fano line-shape of the metamolecule to the effective permeability of the metamaterial. This provides an extra degree of control over the metamaterial's response. An example of this is the ability to tune the frequency at which certain phenomena occur. For example, from Fig.~\ref{fig4F}~(d) one can see that at the MNP-QD resonance point the Fano dip causes a negative real permeability. Thus, by dynamically shifting the detuning between the MNPs and the QDs, one can shift the frequency at which the metamaterial has negative permeability, as shown in Fig.~\ref{fig5}. However, the trade-off for this tunability is that as the QDs are detuned away from the magnetic resonance the bandwidth narrows. The bandwidth, $\delta$, of the dip varies from $0.05$~THz to $0.01$~THz as we detune the QD away from the MNP nanoring's resonance. This still compares favorably to the bandwidth found in EIT experiments with cold rubidium atoms, where a dip bandwidth of $\delta = 50$~MHz is observed~\cite{EITexp}.

\begin{figure}[t]
\begin{center}
\includegraphics[scale=0.44]{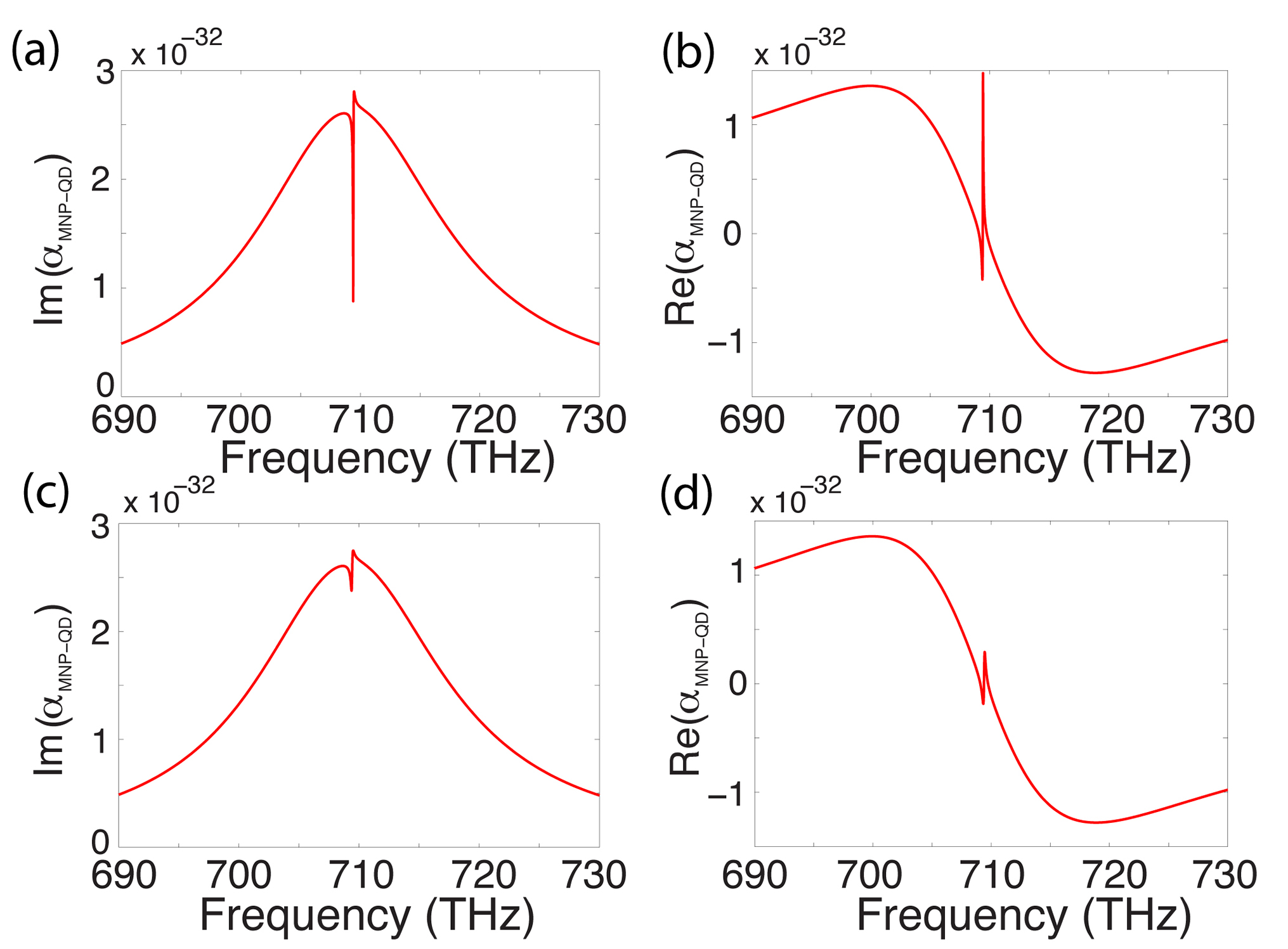}
\caption{\label{fig:5}(Color online) Nonlinear response of the MNP-QD metamolecule. A comparison of the imaginary ((a) and (c)) and real ((b) and (d)) parts of the polarizability for a weak external driving field (top row) with $E_0\mu = 0.0001$ meV, and a strong driving field (bottom row) with $E_0 \mu = 0.1$ meV. The MNP-QD detuning is $\Delta = 0.195 \times 10^{15}$rad s$^{-1}$.}
\label{fig6}
\end{center}
\end{figure}

So far, all calculations have been confined to the weak driving field limit. However if we want to study the nonlinear properties of the nanoring metamaterial we need to consider a strong driving field. If the MNP-QD metamolecule is driven strongly by an intense light field then its optical scattering properties are modified dramatically. The two-level QD becomes saturated by the driving field and its interaction with the MNP field disappears. This nonlinear Fano effect cannot be predicted by classical or semiclassical theory, and has been studied previously in isolated MNP-QD systems and recently observed in quantum-well structures~\cite{NLFano, Qplexcitonics, NLFanoNat}. In Fig.~\ref{fig6} the parameters used in our study are chosen in order to show this effect for a single MNP-QD system of the nanoring. The MNP and the QD are driven by the same external field, and one can see that as the intensity of the field increases (from the top row to the bottom row), the Fano dip in the metamolecule's polarizability is washed out by the saturation of the QDs population. These results have been computed by solving the full master equation numerically. The numerical approach involves solving the eigenproblem, 
\begin{equation}
\hat{\mathcal{L}}(\hat{\rho}_{SS}) = 0,
\end{equation}
where $\hat{\rho}_{SS}$ is the NESS density matrix of the system. The difficulty here lies in the unbounded dimensions of the bosonic MNP field mode, whose Hilbert space is infinite. In order to capture the non-classical behavior of the system the truncation of the dimension of each of the MNP field's Hilbert space has a lower bound of  $d = 15$. This problem is well suited to the quantum optics toolbox developed by Tan~\cite{qotoolbox}.

Our nanoring contains a minimum of four MNPs which makes its combined Hilbert space very large. Thus while the formalism we have developed and studied in this work is now in place, it is unfortunately too computationally intensive at present to study the saturation of the nanoring. However, logically if the QD is saturated in the MNP-QD metamolecule it will also be saturated when coupled to the MNPs in the nanoring. The addition of the QDs into the MNP nanoring cause the material's permeability to become negative at their resonance frequency, as seen in Fig.~\ref{fig4F} and Fig.~\ref{fig5}. Thus, if the QD was saturated by a separate control field then the permeability could be controlled and varied with the light intensity between positive and negative values. In future work, techniques from many-body quantum systems~\cite{manybody} may be used together with our formalism in order to make the computation accessible.


\section{Conclusion}

We have developed a quantum optical model to describe the dynamics of a negative permeability metamaterial integrated with two-level QD systems. Using this model we found that the Fano interference of a MNP-QD metamolecule can manifest itself in the macroscopic magnetic response of a metamaterial consisting of MNP-QD nanorings. We have shown that this effect can be used to tune the properties of the metamaterial. Our model is also useful to study nonlinear effects that arise when the metamaterial is driven strongly. We showed an example of this by studying how the non-linear Fano effect can affect our nanoring.

Each MNP has its own electric response, however the frequencies at which the MNP nanoring metamaterial has a negative permeability and the frequencies at which it has a negative permittivity do not overlap. Even with inclusion of the QDs this problem remains. We have performed calculations which show that the Fano profile in the electric and magnetic scattered fields cannot be tuned independently in our scheme. Without this ability it is not possible to ensure a frequency overlap. Instead, in future work we intend to investigate how to incorporate the magnetic nanoring resonators with a broadband negative permittivity background~\cite{shalaevmetabook}, using various types of lattice configuration. In this way by dynamically tuning the magnetic response, we may also be able to control the metamaterial's refractive index. However, even for a material that has only a negative permeability, recent work has shown interesting quantum dynamics can be observed in the spontaneous emission interference of an emitter placed in close proximity~\cite{negspon}. The ability to tune and saturate the magnetic response in this scenario may open up new additional features. 

Another direction of future work would be to exam how fluorescence quenching of the QD by the MNP would affect our system. In our parameter space we do not expect quenching to be a major factor~\cite{ArtusoQuench}, however it would be interesting to quantify at what point this approximation no longer holds for the system we have studied.
 
Furthermore, in our calculations we have used similar parameters to previous studies~\cite{salandrino, simovskiring,aluring}. However, it has been noted recently that the Ohmic damping used for the MNP fields may well be an underestimation~\cite{moritsPRB2010}, resulting from the discrepancy between the Drude model used theoretically and experimental results at higher frequencies. Using more realistic damping it has been shown that rings with spherical MNPs no longer have a strong enough magnetic resonance to achieve ${\rm Re}(\mu) < 0$. However, this problem may potentially be resolved using MNP's with embedded gain material~\cite{MNPgain} or different type of nanostructures in place of the MNPs. Indeed, by considering more complex, strongly polarized plasmonic nanostructures within the ring~\cite{moritsPRB2010,moritsmeta2011} negative permeability has been shown to be possible, whilst ensuring damping is correctly accounted for. Here, Morits and Simovski have explored the use of dimers~\cite{moritsPRB2010} and nanoprisms~\cite{moritsmeta2011}, with the latter providing negative permeability in the visible regime. In both cases as only dipole interactions are considered, the basic theoretical model we have developed in this work using MNPs can be transferred to these more complex nanostructures in a straightforward manner with the qualitative results of our analysis remaining valid.


\begin{acknowledgments}
This work was supported by the Leverhulme Trust, the UK Engineering and Physical Sciences Research Council and the Qatar National Research Fund (Grant NPRP 4-554-1-D84). We thank T. Tufarelli for useful discussions and V. Giannini for computational resources.
\end{acknowledgments}

\section*{Appendix}
\appendix


\section{THE MNP-QD METAMOLECULE}
\label{appA}

Here we derive expressions for the MNP-QD and MNP-Driving field coupling frequencies, $g$ and $\chi$. We also find an expression for the dipole moment of the MNP used in Section~\ref{metamo}. The coupling frequency, $g$, of the dipole interaction between the MNP field mode and the QD is defined as $\hbar g = \mu\xi$, where $i \xi\hat{a} = \hat{E}_m$ is the positive frequency part of the MNPs dipolar electric field. In order to derive an expression for $g$ and $\chi$, one must equate the NESS quantum expectation value of the MNP electric field with its classically derived value, {\it i.e.} $\langle \hat{E}_m \rangle = E_m$. The classical NESS value is given by 
\begin{equation}
E_m = \frac{S}{4\pi\epsilon_0\epsilon_b}\frac{p_{_{MNP}}}{d^3},\label{classE} \\
\end{equation}
where $p_{_{MNP}}$ is the dipole moment of the MNP, 
\begin{equation}
p_{_{MNP}} = 4\pi\epsilon_0\epsilon_b r^3\bigg(\frac{\epsilon_m(\omega) -\epsilon_b}{\epsilon_m(\omega)+2\epsilon_b}\bigg)\bigg(E_0 + \frac{Sp_{_{QD}}}{4\pi\epsilon_0\epsilon_bd^3}\bigg). \label{A2}
\end{equation}
Here, $S$ is a scalar parameter set to 2 (-1) for the case of the driving field being parallel (perpendicular) to the MNP-QD separation vector, $d$ is the MNP-QD separation distance, $\epsilon_b$ is the permittivity of the background material, $r$ is the radius of the MNP, $E_0$ is the driving field amplitude and $p_{_{QD}}$ is the dipole moment of the QD. From Eq.~(\ref{A2}), we see that the MNP is excited by the external driving field and the QD field. The dipole moment of the QD is given by $p_{_{QD}} = \mu\langle\hat{\sigma}\rangle$. The frequency dependent complex function, $\frac{\epsilon_{m}(\omega) - \epsilon_b}{2\epsilon_b + \epsilon_{m}(\omega)}$, determines the resonance frequency, $\omega_0$, of the MNP field, where $\epsilon_m$ is the permittivity of the metal, calculated with the Drude model. This resonance will occur when the Fr\"ohlich condition is met, ${\rm Re}[\epsilon_{m}(\omega)] = -2\epsilon_b$~\cite{plasmonicsmaier,kreibigclusters}. A first-order Taylor expansion of $\epsilon_m(\omega)$ allows the MNPs polarizability to be approximated by a complex Lorentzian
\begin{align}
\alpha_{MNP} = \frac{12\pi\epsilon_0\epsilon^2_br^3\eta i}{i\Delta_0 +\frac{\gamma_0}{2}},  \label{A4}
\end{align}
where $\gamma_0$ is the total damping rate (radiative and non-radiative) of the MNP. The non-radiative damping, $\gamma_{nr} = \gamma + \frac{\gamma^3(2\epsilon_b+\epsilon_{\infty})}{\omega_p^2}$ comes naturally from $\epsilon_m(\omega)$, while the radiative damping is added in phenomenologically. The parameter $\eta = \frac{(\gamma^2(2\epsilon_b + \epsilon_{\infty}) + \omega_p^2)^2}{2(2\epsilon_b+\epsilon_{\infty})^{\frac{3}{2}}\omega^3_p}$, where $\omega_p$ is the plasma frequency of silver and $\epsilon_{\infty}$ is the ultra-violet permittivity of silver. This approximation allows us to draw an analogy between the plasmonic mode and a leaky cavity mode. The NESS value for the MNPs electric field in the quantum formalism is
\begin{align}
\langle\hat{E}_m\rangle &= i\xi\langle\hat{a}\rangle \\
&=\frac{i\hbar g \langle\hat{a}\rangle}{\mu}.
\end{align}
Substituting Eq.~(\ref{expmnp}) into the above equation we find
\begin{align}
\langle\hat{E}_m\rangle &= \frac{i\hbar g^2\langle\hat{\sigma}\rangle}{\mu(i\Delta_0+ \frac{\gamma_0}{2})} + \frac{-g\chi E_0}{\mu(i\Delta_0 + \frac{\gamma_0}{2})}. \label{A7}
\end{align}
Where $\mu$ is the dipole moment of the QD and $\Delta_{0(x)} = (\omega_{0(x)} - \omega)$. If we substitute Eq.~(\ref{A4}) into Eq.~(\ref{A2}), then the subsequent expression into Eq.~(\ref{classE}) and compare it with Eq.~(\ref{A7}) one obtains the following expressions for $g$ and $\chi$,
\begin{align}
&g = \frac{S\mu}{d^3}\sqrt{\frac{3\eta r^3}{4\pi\epsilon_0\hbar}}, \label{g} \\
&\chi = -i\epsilon_b\sqrt{12\eta\epsilon_0\pi\hbar r^3}. \label{chi}
\end{align}

If we subsequently compare the NESS value of $\langle\hat{a}\rangle$ with the classical expression of $p_{MNP}$ (Eq. (\ref{A2})), we see that
\begin{align}
p_{MNP} = \chi^*\langle\hat{a}\rangle
\end{align}


\section{THE MNP NANORING}
\label{appB}

Here we derive expressions for the nearest neighbor and next-nearest neighbor inter-MNP coupling frequencies, $J_1$ and $J_2$ respectively. In order to calculate the inter-MNP coupling frequencies we take a similar approach as we did in Appendix~\ref{appA}, in that we compare the classical derived values to the expectation values of the quantum formalism. The dipole moment of a single MNP, $p_n$, in an $N=4$ nanoring using classical theory is given by 
\begin{align}
p_{n}  = \frac{12\pi\epsilon_0\epsilon_b^2r^3\eta i}{i\Delta_0 + \frac{\gamma_0}{2}}(E_0+ Q_1p_{n+1} + Q_1p_{n-1} + Q_2p_{n+2}),\label{B1}
\end{align}
where $Q_1$ and $Q_2$ are scalar interaction terms that account for the MNP-MNP couplings in the ring. The above expression takes into account the Lorentzian approximation for the MNP resonance that was used in Appendix~\ref{appA}. Using Eq.~(\ref{aringexp}) the NESS dipole moment in the quantum framework is given by
\begin{align}
\chi^*\langle\hat{a}_n\rangle =& \frac{i|\chi|^2E_0}{\hbar(i\Delta_0 +\frac{\gamma_0}{2})} -\frac{iJ_1\chi^*(\langle\hat{a}_{n+1}\rangle + \langle\hat{a}_{n-1}\rangle)}{i\Delta_0 + \frac{\gamma_0}{2}} \label{B2} \\ \nonumber &-\frac{iJ_2\chi^*\langle\hat{a}_{n+2}\rangle}{i\Delta_0 + \frac{\gamma_0}{2}}.
\end{align}
By comparing Eqs.~(\ref{B1}) and~(\ref{B2}) expressions for $J_1$ and $J_2$ are found to be
\begin{equation}
J_{1(2)} = -12\pi\epsilon_0\epsilon_b^2r^3\eta Q_{1(2)}.
\end{equation}
The inter-QD coupling frequencies, $I_n$ are calculated using the same method.

The value of $Q_{j}$ is found from the general scalar interaction terms $Q_{j\ell}$, where $Q_1=Q_{j\ell}$ for $j$ and $\ell$ nearest neighbors and $Q_2=Q_{j\ell}$ for $j$ and $\ell$ next-nearest neighbors. In the case of the magnetic response excitation, $Q_{j\ell}$ can be defined as the azimuthal component of the electric field at site $j$ due to the azimuthally directed dipole at site $\ell$~\cite{aluring}. The interaction term is therefore calculated from the standard form of a dipolar electric field~\cite{greiner}
\begin{equation}
\bm{E} = \frac{e^{ikr''}}{4\pi\epsilon_{0}\epsilon_{m}}[k^2(\bm{r''} \times \bm{p}) \times \frac{1}{r''^3} + (3\bm{r''}(\bm{p}\cdot\bm{r''}) - \bm{p}r''^2)(\frac{1}{r''^5} - \frac{i}{r''^4})], \label{B4}
\end{equation}
where $\bm{r''}$ is the distance vector between the receiving dipole $j$ at coordinates ($z = 0,R = R_{ring},\phi = \frac{2\pi j}{N}$) and the source dipole $\ell$ at coordinates ($z' = 0, R' = R_{ring},\phi' = \frac{2\pi \ell}{N}$). Using the definition of $Q_{j\ell}$ and Eq.~(\ref{B4}) we derive the expression~\cite{aluring,simovskiring},
\begin{eqnarray}
&&Q_{j\ell} = \frac{e^{ikr''}}{4\pi
\epsilon_{0}\epsilon_{m}r''^5} \label{last} \\ 
&& \qquad \times [(kr'')^2(2R^2 \cos(\phi'-\phi) -R^2 \cos^2(\phi'-\phi)-R^2) \nonumber \\
&& ~~~+ (3R^2 \sin(\phi' - \phi))(1 - ikr'') - (r''^2 \cos(\phi'-\phi))(1-ikr'')]. \nonumber
\end{eqnarray}


\end{document}